\documentclass[final,5p,times]{elsarticle}
\usepackage{graphicx}
\usepackage{dcolumn}
\usepackage{bm}
\usepackage{amsmath}
\usepackage{amssymb}
\usepackage{latexsym}
\usepackage{epsfig}
\usepackage{amsbsy}
\usepackage{array}
\usepackage{amssymb}
\usepackage{setspace}
\usepackage{bm}
\usepackage{epstopdf}
\usepackage{subfigure}

\journal{Journal of \LaTeX\ Templates}

\begin{document}

\begin{frontmatter}
\title{Phonon laser in a cavity magnomechanical system}
\author[mymainaddress]{Ming-Song Ding}
\author[mysecondaryaddress]{Li Zheng}
\author[mymainaddress]{Chong Li\corref{mycorrespondingauthor}}
\cortext[mycorrespondingauthor]{Corresponding author}
\ead{lichong@dlut.edu.com}
%\author[mymainaddress]{Chong Li\corref{mycorrespondingauthor}}
%\ead{lichong@dlut.edu.com}

\address[mymainaddress]{School of Physics, Dalian
University of
Technology, Dalian 116024, China}
\address[mysecondaryaddress]{Information
Science and Engineering College, Dalian Polytechnic University,
Dalian
116034, China}

\begin{abstract}
The phonon analog of an optical laser has become the focus of research. We
theoretically study phonon laser in a cavity magnomechanical system, which
consist of a microwave cavity, a small ferromagnetic sphere and an uniform
external bias magnetic field. This system can realize the phonon-magnon
coupling and the cavity photon-magnon coupling via magnetostrictive
interaction and magnetic dipole interaction respectively, the magnons are
driven directly by a strong microwave field simultaneously. Frist, the
intensity of driving magnetic field which can reach the threshold condition
of phonon laser is given. Then, we demonstrate that the phonon laser can be
well controlled by an adjustable external magnetic field without changing
other parameters, which provides an additional degree of freedom compared to
the phonon laser in optomechanical systems. Finally, with the experimentally
feasible parameters, threshold power in our scheme is close to the case of
optomechanical systems. Our results provide a theoretical basis for the
realization of phonon lasers in magnomechanical systems.
\end{abstract}

\end{frontmatter}

\section{Introduction}

In recent years, cavity magnomechanical system has been
becoming a novel platform for realizing quantum coherence and coupling
between magnons, cavity photons and phonons. Among them, the coupling
between photons and magnons is realized by the magnetic dipole interaction,
and the interaction between magnons and phonons is based on the
magnetostrictive force. As we know the traditional optomechanical systems
utilize radiation force \cite%
{a0,a1,a001,a2,a3,a4,a5,a6,a7,a71,a72,a73,a74,a75,a76,a77}, electrostatic
force \cite{a8,a9}, and piezoelectric force \cite{a10} for coupling phonon
with optical or microwave photons, but they all intrinsically lack good
tunability. The emergence of magnetostrictive force provides us with a new
way to achieve different information carriers \cite{k0,k1}. And the highly
polished single-crystal yttrium iron garnet (YIG) sphere is introduced into
the cavity magnomechanical system as an effective mechanical resonator, the
magnons inside it are collective excitation of magnetization, whose
frequency can be easily adjusted by external bias magnetic field. The
varying magnetization caused by the excitation of the magneton in the YIG
sphere results in the geometric deformation of the surface, introducing the
coupling between magnon and phonon modes.

Due to YIG sphere's high spin density and low damping rate, the Kittel mode
\cite{k15} (the ferromagnetic resonance mode) in it can strongly\cite%
{k150,k3,k5} to the microwave cavity photons $(g_{ma}>\kappa _{a}$, $\kappa
_{m})$ in cavity-magnon systems. In addition, the YIG sphere has rich
magnonic nonlinearities and the characteristic of low loss in different
information carriers, these excellent properties make it possible to find
many interesting and important phenomena in cavity-magnon systems and cavity
magnomechanical systems. Based on it, a lot of theoretical and experimental
researches have been done. J. Q. You $et$ $al$. have found the bistability
of cavity magnon polaritons \cite{k6}, G. S. Agarwal $et$ $al$. have
discussed the tripartite entanglement among magnons, cavity photons, and
phonons\cite{k0}. Furthermore, high-order sideband generation \cite{k21,k22}%
, magnon Kerr effect \cite{k2}, the light transmission in cavity-magnon
system \cite{k23} and other researches were also studied \cite%
{k24,k25,k26,k3,k5,k7,k70,k151}.

Phonon laser as a novel laser has been developed rapidly, it generates
coherent sound oscillations (mechanical vibration) by optical pumping. Just
like traditional optical laser \cite{k71,k72}, phonon laser can be
considered as an analogue of a two-level optical laser which is provided by
phonon mediated transitions between two optical supermodes \cite{k8,k9,k10}.
These supermodes correspond to the ground and excited states respectively,
and the mechanical mode (phonons) mediates the transition between them. As
early as 2003, Chen. J and Khurgin have verified the feasibility of phonon
lasers and proposed a scheme to realize phonon lasers \cite{k101}. Then
single trapped ions and quantum dots have been utilized in the fields of
phonon laser \cite{k102,k103,k104}. Up to now, numerous theoretical and
experimental researches have been proposed, such like the cavity
optomechanics-based ultralow-threshold phonon lasers \cite{k8}, the $%
\mathcal{PT}$-symmetric phonon laser with balanced gain and loss \cite{k9},
the nonreciprocal phonon lasing in a coupled cavity system \cite{k10}, the
phonon laser operating at an exceptional point \cite{k105}, the scheme of
amplifying phonon laser by using phonon stimulated emission coherence \cite%
{k106}, the phonon-stimulated emission in cryogenic ionic compounds \cite%
{k107,k108}, semiconductor superlattices \cite{k110} and so on \cite%
{k109,k111,k113,k114}. In addition, phonon laser have also attracted
extensive interest in medical imaging and high-precision measurement
equipment.

In this work, we study a cavity magnomechanical system, which consist of a
microwave cavity, a YIG sphere and the uniform external bias magnetic field $%
H$ (vertical direction). The magnetostrictive (radiation pressure like)
interaction mediates the coupling between magnons and phonons, and the
photons and magnons are coupled via magnetic dipole interaction. It is worth
noting that unlike optical pump in the traditional cavity optomechanical
system, we introduce magnetic driving field to to realize phonon laser.
Furthermore, the magnomechanical interaction which is quite weak in
experiments can be enhanced by the gain of magnon mode. We found that the
production of laser can be well modulated by adjusting the applied magnetic
field $H$, which provides an additional degree of freedom to control phonon
laser action. It is worth mentioning that the applied magnetic field $H$,
the drive magnetic field, and the magnetic field of the cavity mode are
mutually perpendicular at the site of the YIG sphere. So we can adjust only
one of them without worrying about the impact on the rest. Then, the
threshold conditions of driving magnetic field intensity for phonon laser is
given. We can make our system reach the threshold condition by enhancing
drive magnetic field. And the threshold power required can be below $10\mu $%
W within the experimental allowable range of parameters. According to the
recent work, the threshold power in cavity optomechanical system is
generally about $7\mu $W \cite{k8,k9,k10}. Accordingly, phonon laser has
potential application value in cavity magnomechanical systems.

\section{Model and dynamical equations}
\begin{figure}[tbp]
\label{fig.2} %,height=0.40\columnwidth
\centering
\includegraphics*[width=0.74\columnwidth]{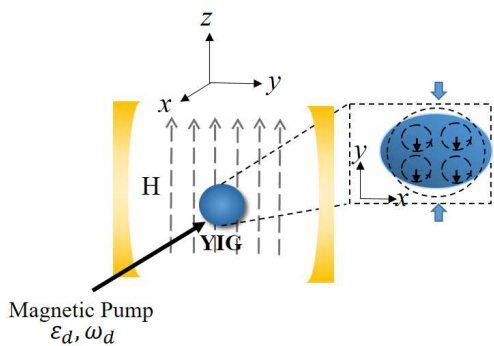}
\caption{(1)Schematic illustration of the system, a YIG sphere is placed in
the maximum magnetic field of a microwave cavity mode. And an uniform
external bias magnetic field H is applied along the z-direction to bias the
YIG sphere. The enlarged YIG sphere on the right illustrates how the dynamic
magnetization of magnon (vertical black arrows) causes the deformation
(compression along the y-direction) of the YIG sphere (and vice versa),
which rotates at the magnon frequency. Furthermore, a microwave source is
used to drive the magnon mode. It's important to note that the bias magnetic
field (z direction), the drive magnetic field (y direction), and the
magnetic field (x direction) of the cavity mode are mutually perpendicular
at the site of the YIG sphere. }
\end{figure}
We consider a hybrid cavity
magnomechanical system, which consists a microwave cavity and a small sphere
(a highly polished single-crystal YIG sphere of diameter $1mm$ is used in
\cite{k2}). There are three modes in this system: cavity photon mode, magnon
mode and phonon mode. As shown in Fig. 1, the YIG sphere is placed near the
maximum microwave magnetic field of the cavity mode, an uniform external
bias magnetic field $H$ is applied along the $z$ direction to bias the YIG
sphere simultaneously, which establish the magnon-photon coupling \cite%
{k2,k6}. The magnon-photon coupling can be tuned by moving the YIG sphere
inside the cavity. In addition, the magnetic field $H$ is created by a high
precision tunable electromagnet, and the adjusting range of bias magnetic
field $H$ is between $0$ and $1T$ \cite{k2}.

Here, the coupling between magnons and phonons is generated by
magnetostrictive interaction (the derivation of the relevant Hamiltonian can
be found in \cite{k1}). Beause of the varying magnetization induced by the
magnon excitation inside the YIG sphere, the sphere\ produce micro
deformation and it can be used as an excellent mechanical resonator. Based
on it, we have the vibrational modes (phonons) of the sphere. Here, a
microwave source is used to directly drive the magnon mode and it can
enhance the magnomechanical coupling \cite{k2,k0}. It is worth mentioning
that the applied magnetic field $H$, the drive magnetic field, and the
magnetic field of the cavity mode are mutually perpendicular at the site of
the YIG sphere. So we can adjust only one of them without worrying about the
impact on the rest. Furthermore, we concurrently assume that the size of the
sphere is so smaller than the wavelength that the interaction between cavity
microwave photons and phonons can be neglected. The total Hamiltonian of the
hybrid system reads ($\hbar =1$)

\begin{eqnarray}
H_{total} &=&H_{0}+H_{int}+H_{d},  \nonumber \\
H_{0} &=&\omega _{a}a^{\dagger }a+\omega _{m}m^{\dagger }m+\omega
_{b}b^{\dagger }b,  \label{eq1} \\
H_{int} &=&g_{ma}(a^{\dagger }m+m^{\dagger }a)-g_{mb}m^{\dagger
}m(b+b^{\dagger }),  \nonumber \\
H_{d} &=&i(\varepsilon _{d}m^{\dagger }e^{-i\omega _{d}t}-\varepsilon
_{d}^{\ast }me^{i\omega _{d}t}),  \nonumber
\end{eqnarray}%
where $H_{0}$ is the free Hamiltonian, the first and second terms denote the
cavity photon mode and magnon mode, respectively. The third term describes
the mechanical mode. $\omega _{a}$, $\omega _{m}$ and $\omega _{b}$ denote
the resonance frequencies of the cavity, magnon, and mechanical modes. Here,
$\omega _{m}$ is the frequency the Kittel mode. An uniform magnon mode
resonates in the YIG sphere at frequency $\omega _{m}=\gamma _{g}H$, where $%
\gamma _{g}$ is gyromagnetic ratio and $\gamma _{g}/2\pi =28GHz/T$. The
annihilation (creation) operators of these modes are $a(a^{\dagger })$, $%
m(m^{\dagger })$ and $b^{\dagger }(b)$, respectively.

$H_{int}$ is the interaction Hamiltonian of system, the first term of $%
H_{int}$ is the coupling between the cavity and magnon modes, the second
term represents phonon-magnon interaction. $g_{ma}$ and $g_{mb}$ are the
coupling rates of the magnon-cavity interaction and the magnon-phonon
interaction, respectively. And we can tune $g_{ma}$ by adjusting the
direction of bias field or the position of the YIG sphere inside the cavity.
Finally, $H_{d}$ is the Hamiltonian which describes the external driving of
the magnon mode, as shown in \cite{k2}, J. Q. You $et$ $al$ designed an
experimental setup, the YIG sphere can be directly driven by a
superconducting microwave line which is connected to the external port of
the cavity. Rabi frequency $\varepsilon _{d}=\frac{\sqrt{5}}{4}\gamma _{g}%
\sqrt{M}B_{0}$ (under the assumption of the low-lying excitations) stands
for the coupling strength of the drive magnetic field \cite{k0}, the
amplitude and frequency are $B_{0}$ and $\omega _{d}$ respectively. The
total number of spins $M=\rho V$, where $V$ is the volume of the sphere.
Furthermore, $\rho =$ $4.22\times 10^{27}m^{-3}$ is the spin density of the
YIG sphere.

By making a frame rotating at the drive frequency $\omega _{d}$ and using
rotating-wave approximation, the total Hamiltonian of the system can be
rewritten as

\begin{eqnarray}
H_{total} &=&-\Delta _{a}a^{\dagger }a-\Delta _{m}m^{\dagger }m+\omega
_{b}b^{\dagger }b+  \nonumber \\
&&g_{ma}(a^{\dagger }m+m^{\dagger }a)-g_{mb}m^{\dagger }m(b+b^{\dagger })
\label{eq2} \\
&&+i(\varepsilon _{d}m^{\dagger }-\varepsilon _{d}^{\ast }m),  \nonumber
\end{eqnarray}%
where $\Delta _{a}=$ $\omega _{d}-\omega _{a}$ is the detuning between the
driving field and cavity mode, $\Delta _{m}=$ $\omega _{d}-\omega _{m}$
denotes the detuning between the driving field and resonance frequency of
cavity mode. The Heisenberg-Langevin equations of the system are given by

\begin{eqnarray}
\dot{a} &=&(i\Delta _{a}-\kappa _{a})a-ig_{ma}m-\sqrt{2\kappa _{a}}a_{int},
\nonumber \\
\dot{m} &=&(i\Delta _{m}-\kappa _{m})m-ig_{ma}a+ig_{mb}m(b+b^{\dagger })
\label{eq3} \\
&&+\varepsilon _{d}-\sqrt{2\kappa _{m}}m_{int},  \nonumber \\
\dot{b} &=&(-i\omega _{b}-\gamma _{b})b+ig_{mb}m^{\dagger }m-\xi _{no},
\nonumber
\end{eqnarray}%
where $\gamma _{b}$ is the mechanical decay rate, $a_{int},m_{int}$ and $\xi
_{no}$ are input noise operators of cavity, magnon and mechanical modes
respectively. $\kappa _{m}$ and $\kappa _{a}$ are the losses of magnon and
microwave cavity modes. Like the computation process of phonon laser \cite%
{k8,k10}, assuming that the magnon mode is strongly driven, leading to a
large amplitude $\left\vert \left\langle m\right\rangle \right\vert $ $\gg 1$
at the steady state, and due to the cavity-magnon beam splitter interaction,
the cavity field also has a large amplitude $\left\vert \left\langle
a\right\rangle \right\vert $ $\gg 1$. That leads to the quantum noise terms
can be safely neglected if one is interested only in the mean-number
behaviors (i.e., the threshold feature of the mechanical gain or the phonon
amplifications). Therefore, the semi-classical Langevin equations of motion
are used. In other words, we can rewritten all operators as their respective
expectation values. Then by setting the left-hand side equal to zero, the
steady-state mean values of the system read

\begin{eqnarray}
a_{s} &=&\frac{g_{ma}\cdot m_{s}}{\Delta _{a}-i\kappa _{a}},  \nonumber \\
m_{s} &=&\frac{\varepsilon _{d}}{(\kappa _{m}-\frac{g_{ma}^{2}\Delta _{a}}{%
\Delta _{a}^{2}-\kappa _{a}^{2}})-i[\Delta _{m}+g_{mb}(b_{s}+b_{s}^{\ast })+%
\frac{g_{ma}^{2}\kappa _{a}}{\Delta _{a}^{2}-\kappa _{a}^{2}}]},  \label{eq5}
\\
b_{s} &=&\frac{g_{mb}\left\vert m_{s}\right\vert ^{2}}{\omega _{b}-i\gamma
_{b}}.  \nonumber
\end{eqnarray}

According to the feasible experimental parameters ($g_{mb}<1Hz$), $%
g_{mb}(b_{s}+b_{s}^{\ast })\ll $ $\Delta _{m}$. Under this condition, we
approximately have $\Delta _{m}+g_{mb}(b_{s}+b_{s}^{\ast })\sim \Delta _{m}$.
In close analogy to an
optical laser, a coherent emission of phonons can be achieved with two
coupled whispering-gallery-mode microtoroid resonators via inversion of the
two optical supermodes. This leads to phenomenon of phonon laser at the
breathing mode, with the threshold power $P_{th}\sim 7\mu $W \cite{k8,k12}.
Similarly, our system also has two supermodes corresponding to the ground
and excited states of the two-level system, respectively. The mechanical
mode (phonon) can realize energy level transition between levels, the
stimulated emission of phonon can be generated by virtue of magnetic pumping
of the upper level, then leading to the appearance of coherent phonon
lasing. Therefore we introduce supermode operators $\Re _{\pm }=(a\pm
m^{\dagger })/\sqrt{2}$ to rewrite the Hamiltonian $H_{0}$ and $H_{d}$ of
the system, i.e.,

\begin{eqnarray}
H_{0,sm} &=&\omega _{+}\Re _{+}^{\dagger }\Re _{+}+\omega _{-}\Re
_{-}^{\dagger }\Re _{-}+\omega _{b}b^{\dagger }b,  \label{eq6} \\
H_{d,sm} &=&i/\sqrt{2}[\varepsilon _{d}(\Re _{+}^{\dagger }+\Re
_{-}^{\dagger })-\varepsilon _{d}^{\ast }(\Re _{+}+\Re _{-})],  \nonumber
\end{eqnarray}%
where the supermode frequencies $\omega _{\pm }=-\frac{\Delta }{2}\pm g_{ma}$%
. $H_{int}$ in Eq.(\ref{eq1}) can be tranformed to

\begin{equation}
H_{int}=-\frac{g_{mb}}{2}[(n_{+}+n_{-})+(\Re _{+}^{\dagger }\Re _{-}+\Re
_{+}\Re _{-}^{\dagger })](b+b^{\dagger }),  \label{eq7}
\end{equation}%
with $n_{+}=\Re _{+}^{\dagger }\Re _{+}$ and $n_{-}=\Re _{-}^{\dagger }\Re
_{-}$. In the frame rotating with respect to $H_{0,sm}$ and applying the
rotating-wave approximation, $H_{int}$ is rewritten as

\begin{equation}
H_{int,sm}=-\frac{g_{mb}}{2}(p^{\dagger }b+pb^{\dagger }),  \label{eq8}
\end{equation}%
where $\hat{p}=\Re _{+}\Re _{-}^{\dagger }$ is ladder operator. Eq.(\ref{eq8}%
) represents the absorption and emission of phonons. In general, the
introduction of supermode operators $\Re _{\pm }$ means that the magnon mode
and the optical mode have the same resonant frequency. After changing the
Hamiltonian into the supermode picture, the equations of motion read

\begin{eqnarray}
\dot{\Re}_{+} &=&-(i\omega _{+}+\gamma )\Re _{+}+\frac{i}{2}g_{mb}b\Re _{-}+%
\frac{\varepsilon _{d}}{\sqrt{2}},  \nonumber \\
\dot{\Re}_{-} &=&-(i\omega _{-}+\gamma )\Re _{-}+\frac{i}{2}g_{mb}b^{\dagger
}\Re _{+}+\frac{\varepsilon _{d}}{\sqrt{2}},  \label{eq9} \\
\dot{b} &=&-(i\omega _{b}+\gamma _{b})b+\frac{i}{2}g_{mb}p,  \nonumber \\
\dot{p} &=&-2(\gamma +ig_{ma})p-\frac{i}{2}g_{mb}b\Delta n+\frac{1}{\sqrt{2}}%
(\varepsilon _{d}\Re _{-}^{\dagger }+\varepsilon _{d}^{\ast }\Re _{+}),
\nonumber
\end{eqnarray}%
where $\gamma =(\kappa _{a}+\kappa _{m})/2$, and $\Delta n=n_{+}-n_{-}$ is
inversion operator. Then we set the left-hand side of Eq.(\ref{eq9}) equal
to zero, the zero-order steady states of the system are given by

\begin{eqnarray}
\Re _{+,s} &=&\frac{\sqrt{2}\varepsilon _{d}[2r+i(2\omega _{-}+bg_{mb})]}{%
4(\gamma ^{2}+g_{ma}^{2})-\Delta ^{2}+g_{mb}^{2}b^{\dagger }b-4i\gamma
\Delta },  \nonumber \\
\Re _{-,s} &=&\frac{\sqrt{2}\varepsilon _{d}[2r+i(2\omega _{+}+b^{\dagger
}g_{mb})]}{4(\gamma ^{2}+g_{ma}^{2})-\Delta ^{2}+g_{mb}^{2}b^{\dagger
}b-4i\gamma \Delta },  \label{eq10} \\
p &=&\frac{\sqrt{2}((\varepsilon _{d}\hat{a}_{-}^{\dagger }+\varepsilon
_{d}^{\ast }a_{+})-ig_{mb}b\Delta n}{4\gamma +i(4g_{ma}-2\omega _{b})},
\nonumber
\end{eqnarray}%
where $\Delta =\Delta _{m}+\Delta _{a}$, then Eq.(\ref{eq10}) is substituted
into the dynamical equation of $b$ in Eq.(\ref{eq9}), the result can be
obtained

\begin{equation}
\dot{b}=\Delta _{b}b+\chi ,  \label{eq11}
\end{equation}%
with
\begin{eqnarray}
\Delta _{b} &=&-i\omega _{b}^{^{\prime }}-G-\gamma _{b},  \nonumber \\
G &=&g_{mb}^{2}\gamma \lbrack \frac{\Delta n}{8\gamma ^{2}+2\eta ^{2}}+\beta
],  \label{eq12} \\
\beta &\simeq &\frac{\left\vert \varepsilon _{d}\right\vert ^{2}\eta \Delta
}{4(\gamma ^{2}+g_{ma}^{2}-\frac{\Delta ^{2}}{4}+\Delta ^{2}\gamma
^{2})(\eta +4\gamma ^{2})},  \nonumber
\end{eqnarray}%
where $\eta =2g_{ma}-\omega _{b}$, and the approximation is due to the
parameters we have chosen $(g_{mb}\ll $ $\Delta )$. Then the inversion
operator can be express as

\begin{equation}
\Delta n\simeq \frac{2g_{ma}\left\vert \varepsilon _{d}\right\vert ^{2}}{%
(\gamma ^{2}+g_{ma}^{2}-\frac{\Delta ^{2}}{4})^{2}+\gamma ^{2}\Delta ^{2}}.
\label{eq13}
\end{equation}
\begin{figure}[tbp]
\centering
\includegraphics*[width=0.74\columnwidth]{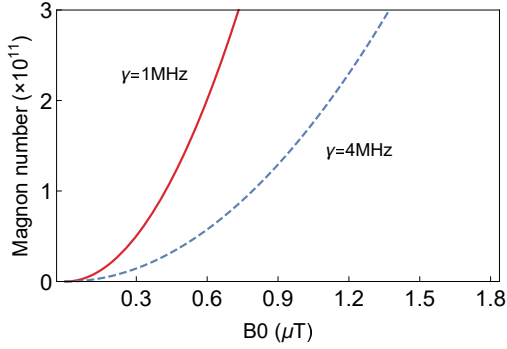}
\caption{The distribution of steady-state magnon number $%
\left\vert m_{s}\right\vert ^{2}$ versus the drive magnetic field $B_{0}$
under $\gamma =1MHz$ (red solid line) and $\gamma =4MHz$ (blue dashed line).
The parameter we used is $\Delta _{m}/2\pi =8MHz$.}
\label{fig:2}
\end{figure}
Because this paper mainly studies the phonon laser generated by the system,
we are only interested in $G$, which indicate the mechanical gain of system.
Therefore, only the specific expression of $G$ is given. The non-negative
mechanical gain $G$ decreases the effective damping rate of the mechanical
mode $\gamma _{eff}=\gamma _{b}-G$, that leads to the instabilities of the
mechanical oscillator at $\gamma _{eff}<0$. This problem has been analyzed
and discussed in \cite{k8,k10,k11} from both theoretical and experimental
perspectives.
\section{The distribution of steady-state magnon number}
Here, we give the
specific values of the parameters used in this paper \cite{k1}. $\omega
_{a}/2\pi =\omega _{m}/2\pi =10.1GHz$, $\omega _{b}/2\pi =12MHz$, $%
g_{ma}/2\pi =6MHz$, $g_{mb}/2\pi =0.1Hz$, $\Delta _{a}/2\pi =8MHz$, and the
loss of mechanical modes $\gamma _{b}/2\pi =100Hz$. Our research is in
resolved sideband regime ($\kappa _{m}/\omega _{b}<1$). The drive power $%
P=(B_{0}^{2}/2\mu _{0})Ac$ \cite{k0}, where $B_{0}^{2}/2\mu _{0}$ is time
average of energy per unit volume, $c$ is the speed of an electromagnetic
wave propagating through the vacuum and $A$ is the maximum cross-sectional
area of YIG sphere. Fig. 2 shows the distribution of steady-state magnon
number $\left\vert m_{s}\right\vert ^{2}$ versus the drive magnetic field $%
B_{0}$. The number of magnons increases exponentially with the increase of $%
B_{0}$, which represents significant nonlinearity. The corresponding $B_{0}$
is much weak relative to the external magnetic field $H$. In addition, the
results under different losses of supermode $\gamma $ are also given. It can
be seen that the smaller the dissipation, the faster $\left\vert
m_{s}\right\vert ^{2}$ increases.

\section{Magnetic field-based control of phonon laser action}

In order to
explore the relationship between the generation of phonon laser and magnetic
field (including the external bias magnetic field $H$ and the drive magnetic
field $B_{0}$), we investigate the phonon number as a function of $H$ and $%
B_{0}$. The mechanical gain $G$ has been given in Eq.(\ref{eq12}), thus the
stimulated emitted phonon number can be calculated \cite{k9,k10}, i.e.,

\begin{equation}
N_{b}=\exp [2(G-\gamma _{b})/\gamma _{b}],  \label{eq14}
\end{equation}%
then from the above expression, the threshold condition of phonon laser is
given (the threshold condition for phonon lasing $N_{b}=1$). When $N_{b}=1$,
we have

\begin{equation}
B_{0,th}=\frac{8\sqrt{\frac{2}{5}\gamma _{b}\Gamma }}{g_{mb}\sqrt{%
g_{ma}M\gamma \Delta }},  \label{eq15}
\end{equation}%
where $\Gamma =g_{ma}^{4}+2g_{ma}^{2}(\gamma ^{2}-\frac{\Delta ^{2}}{4}%
)+(\gamma ^{2}+\frac{\Delta ^{2}}{4})^{2}$. $B_{0,th}$ is the driving
magnetic field required to achieve the threshold condition of the phonon
laser in our system. Finally, according to the expression given earlier $%
P=(B_{0}^{2}/2\mu _{0})Ac$, the threshold power is defined as

\begin{equation}
P_{th}=\frac{64}{5}\frac{Ac\gamma _{b}\Gamma }{g_{mb}^{2}g_{ma}M\gamma \mu
_{0}\Delta }.  \label{eq16}
\end{equation}
\begin{figure}[tbp]
\centering
\includegraphics*[width=0.74\columnwidth]{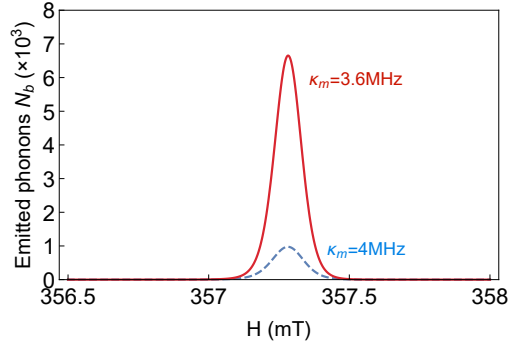}
\caption{The stimulated emitted phonon number $n_{b}$ versus the external
bias magnetic field $H$ under $\protect\kappa _{m}=3.6MHz$ (red solid line)
and $\protect\kappa _{m}=4MHz$ (blue dashed line). The parameter we used is $%
\protect\varepsilon _{d}=1.45\times 10^{11}Hz$ ($B_{0}=2\protect\mu T,P=23%
\protect\mu W$) and $\protect\kappa _{a}=3MHz$.}
\end{figure}
In Fig. 3, $N_{b}$ is plotted as a function of the external bias magnetic
field $H$. There is an obvious window between $H\approx 357mT-357.5mT$ which
generates a large number of the stimulated emitted phonon. This behavior is
consistent with previous effect on the number of magnons, and the reason for
it is the cavity-magnon beam splitter interaction. We find that the
distribution of the stimulated emitted phonon number has a Lorentzian-like
shape dependence on the applied magnetic field $H$, which presents an
additional degree of freedom to control phonon laser. And that phenomenon
similar to a switch of a phonon laser can be obtained by adjusting $H$
without changing other parameters. Especially, the widths of the windows
almost keep unchanged with the loss of magnon mode $\kappa _{m}$. Moreover,
it can be seen that the number of phonons increases with decreasing $\kappa
_{m}$.

In Fig. 4, $N_{b}$ is plotted as a function of the drive magnetic field $%
B_{0}$, the stimulated emitted phonon number is enhanced by input driving
magnetic field. Note that the threshold condition of phonon laser can be
reached at $N_{b}\geqslant 1$. Then we find the loss of supermode has
significant impact on the threshold power $P_{th}$. For $\gamma =1MHz$ and $%
\gamma =4MHz$, we have $P_{th}\sim 0.57\mu $W and $P_{th}\sim 9.79\mu $W,
respectively. It is worth mentioning that $P_{th}$ is calculated in the
range of parameters that can be achieved by experiments, and it is not much
different from the $P_{th}$ obtained in the cavity optomechanics system, and
the threshold power in cavity optomechanical system is generally about $7\mu
$W in \cite{k8,k9,k10,k13}.

\begin{figure}[tbp]
\centering
\includegraphics*[width=0.74\columnwidth]{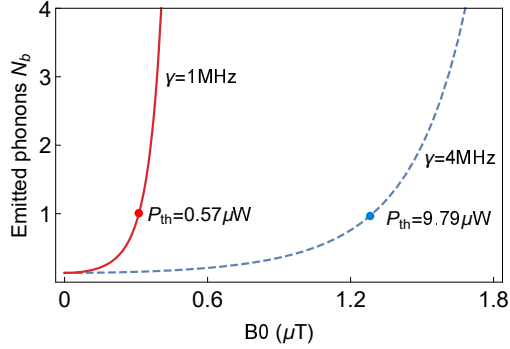}
\caption{The stimulated emitted phonon number $n_{b}$ versus the drive
magnetic field $B_{0}$ under $\protect\kappa _{m}=\protect\kappa _{a}=1MHz$
(red solid line, $P_{th}=0.57\protect\mu $W) and $\protect\kappa _{m}=%
\protect\kappa _{a}=4MHz$ (blue dashed line, $P_{th}=9.79\protect\mu $W).
The other parameter we used is $\Delta _{m}/2\protect\pi =8MHz.$}
\end{figure}

\section{Conclusion}

In summary, we have investigated theoretically phonon laser in a hybrid
cavity magnomechanical system, which use magnetostrictive force (radiation
pressure like) to achieve interaction between magnon mode and mechanical
mode. Our results have shown that by adjusting the the external bias
magnetic field $H$, a window which generates a large number of phonons can
be obtained. And the width of this window is about $0.5mT$. Compared with
the conventional phonon laser work \cite{k8,k9,k10}, our scheme provides an
additional degree of freedom to control phonon laser action.

We frist consider a novel drive magnetic field, which can be realized by
directly driving the YIG sphere with a microwave source. The threshold
conditions of driving magnetic field intensity for phonon laser is given.
Then, we find the system can reach the threshold power and produce phonon
laser by increasing the drive magnetic field $B_{0}$. With the
experimentally feasible parameters, threshold power $P_{th}$ in our system
is close to the threshold power of phonon laser in optomechanical systems
which are mature in theory and experiments. Finally, we hope that the phonon
laser in the hybrid cavity magnomechanical system will be accessible in the
near future.

\section{acknowledgments}

This work was supported by the National Natural Science Foundation of China,
under Grant No. 11574041 and No. 11475037.

\end{document}